\documentclass[preprint,12pt]{elsarticle}
\usepackage{tabularx}
\usepackage{multirow}
\usepackage{makecell}
\usepackage{booktabs}
\usepackage{tabularray}

\usepackage{amssymb}

\journal{Computers in Biology and Medicine}

\begin{document}

\begin{frontmatter}



\title{Artificial intelligence techniques in inherited retinal diseases: A review}


\author[inst1]{Han Trinh}

\affiliation[inst1]{organization={Department of Optometry, School of Allied Health, The University of Western Australia},
            addressline={39 Fairway}, 
            city={Crawley},
            postcode={6009}, 
            state={Western Australia},
            country={Australia}}

\author[inst2]{Jordan Vice}
\author[inst1]{Jason Charng}
\author[inst1]{Zahra Tajbakhsh}
\author[inst1]{Khyber Alam}
\author[inst3]{Fred K. Chen}
\author[inst2]{Ajmal Mian}

\affiliation[inst2]{organization={School of Physics, Maths and Computing, Computer Science and Software Engineering, The University of Western Australia},
            addressline={35 Stirling Highway}, 
            city={Perth},
            postcode={6009}, 
            state={Western Australia},
            country={Australia}}

\affiliation[inst3]{organization={Centre for Ophthalmology and Visual Sciences (Lions Eye Institute), The University of Western Australia},
            addressline={2 Verdun Street}, 
            city={Nedlands},
            postcode={6009}, 
            state={Western Australia},
            country={Australia}}

\begin{abstract}
Inherited retinal diseases (IRDs) are a diverse group of genetic disorders that lead to progressive vision loss and are a major cause of blindness in working-age adults. The complexity and heterogeneity of IRDs pose significant challenges in diagnosis, prognosis, and management. Recent advancements in artificial intelligence (AI) offer promising solutions to these challenges. However, the rapid development of AI techniques and their varied applications have led to fragmented knowledge in this field. 
This review consolidates existing studies, identifies gaps, and provides an overview of AI's potential in diagnosing and managing IRDs. It aims to structure pathways for advancing clinical applications by exploring AI techniques like machine learning and deep learning, particularly in disease detection, progression prediction, and personalized treatment planning. Special focus is placed on the effectiveness of convolutional neural networks in these areas.
Additionally, the integration of explainable AI is discussed, emphasizing its importance in clinical settings to improve transparency and trust in AI-based systems. The review addresses the need to bridge existing gaps in focused studies on AI's role in IRDs, offering a structured analysis of current AI techniques and outlining future research directions. It concludes with an overview of the challenges and opportunities in deploying AI for IRDs, highlighting the need for interdisciplinary collaboration and the continuous development of robust, interpretable AI models to advance clinical applications.
\end{abstract}




\end{frontmatter}


\section{Introduction}
\begin{figure}[h]
    \centering
    \includegraphics[width=1\linewidth]{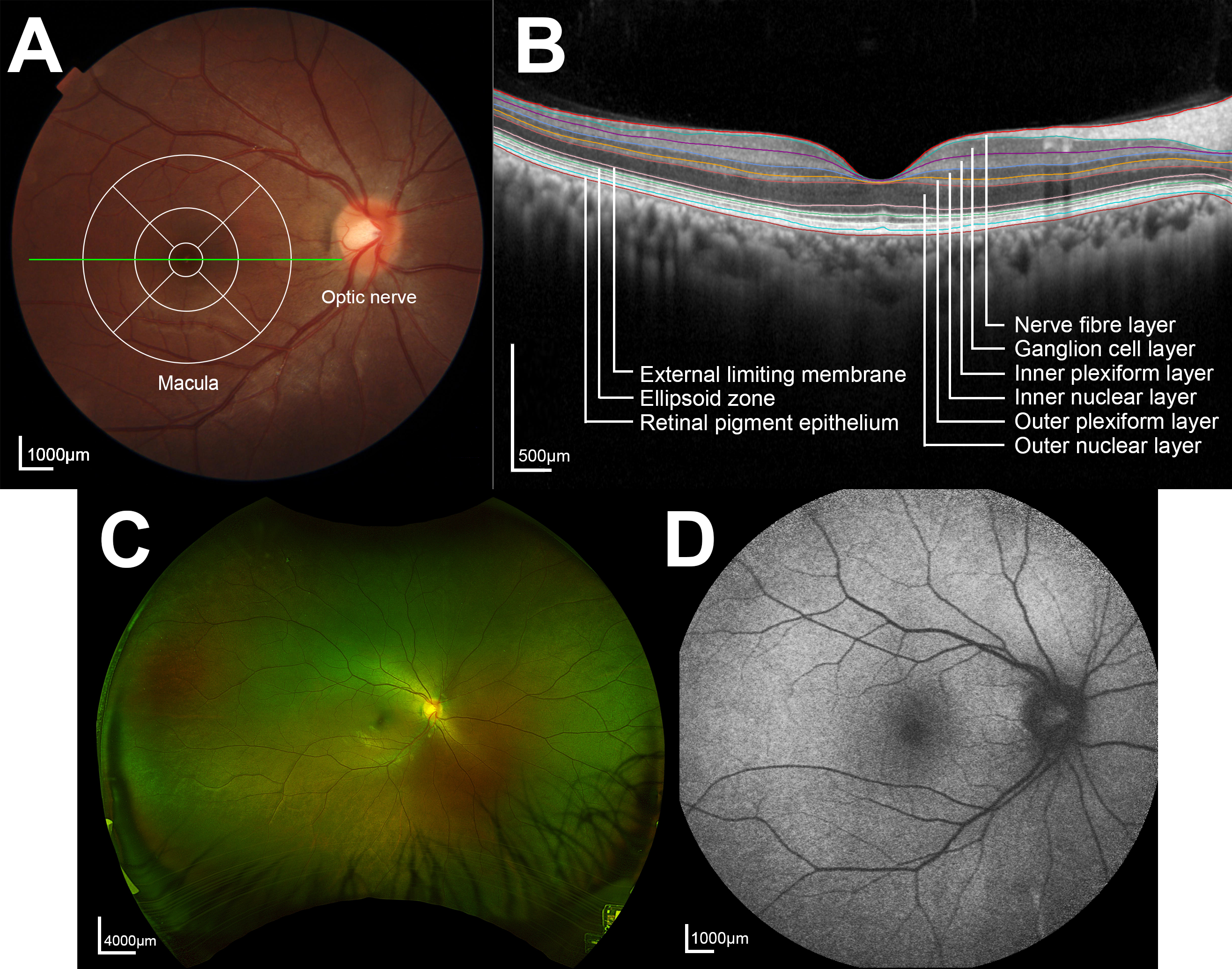}
    \caption{(A) Fundus photo demonstrating an en-face view of the optic nerve and macula. (B) Optical coherence tomography scan showing a cross-section of the macula (through green horizontal line in Figure 1A). (C) Wide-field image of the retina. (D) Short-wave autofluorescence image of the fundus.}
    \label{compilation.png}
\end{figure}

Vision 
allows us to interpret and interact with our surroundings. Vision impairment affects 216.6 million people worldwide \cite{Flaxman2017}, and is associated with significant economic burden \cite{Koberlein2013}, as well as higher mortality and rates of depression and anxiety compared with normally-sighted people \cite{Demmin2020, Ehrlich2021}.
\\
The retina is a critical structure in the eye that plays a key role in vision. It converts light into electrical signals, which are transmitted via the optic nerve to the brain for processing. Composed of multiple cell layers (Figure \ref{compilation.png}), the retina's health is essential for functional vision. Any diseases or conditions affecting the retina can lead to impaired vision.
\\
\\
Inherited retinal diseases (IRDs) are a group of heterogenous genetic ocular conditions that cause vision impairment due to degeneration of retinal layers leading to dysfunction or abnormal development. IRDs are estimated to affect over 5.5 million people worldwide (0.072\%) \cite{Hanany2020}, and is the leading cause of blindness in working age adults \cite{Heath2021}. 
\\
IRDs are caused by pathogenic variants in nuclear and mitochondrial genes, with over 250 genes implicated in this heterogeneous group of diseases \cite{RetNet}. This genetic diversity leads to varied pathophysiology and affects different retinal layers. Even among individuals with the same mutation, disease onset and severity can vary, making it challenging for clinicians to manage uncertainties in disease progression. Retinitis pigmentosa (RP) and Stargardt disease are among the most commonly studied IRDs, with approximately 29.6\% and 24.8\% of IRD-related genes associated with these conditions respectively \cite{Schneider2022}.
\\
In optometry and ophthalmology clinics, patients are assessed for IRDs using a wide range of methods. Structural measures include photographs of the retina, known as fundus photographs, which provide en face views across the entire structure. Fundus autofluorescence (FAF) uses short wavelength and/or near-infrared excitation stimuli to produce en face images of the retinal pigment epithelium (RPE), a layer of the retina responsible for metabolism. This allows for identification of retinal areas of hyperautofluorescence associated with accumulation of lipofuscin and melanin\cite{Keilhauer2006, Smith2009}, which indicate harm to the retina, and areas of hypoautofluorescence associated with atrophy \cite{Smith2009}, which indicates death of retinal cells.  Optical coherence tomography (OCT) uses low coherence interferometry to generate cross sections of the retina \cite{Huang1991}, allowing practitioners to determine which retinal layers are affected. Adaptive optics scanning laser ophthalmoscopy (AOSLO) can also be utilized to visualize the structure of the fundus in high resolution and minimise wavefront distortions \cite{Roorda2002}, however the technique is not readily available in primary eye care settings.
\\
From a functional assessment perspective, visual acuity (VA) is measured to determine high contrast resolution at various distances, and visual field (VF) testing, or perimetry, is conducted to measure the extent of a patient’s peripheral light sensitivity and both of these can be tracked over time \cite{Fishman2007, Grover1998}. Fundus-tracked perimetry is a variation of VF testing that ensures precise stimulation of specific retinal areas by aligning stimulus presentation with a fundus image \cite{Josan2021}. Electroretinography (ERG) measures electrical activity from the retina in response to light stimuli. The retina comprises multiple cell types, and ERG allows practitioners to determine which subset/s of retinal cells are dysfunctional \cite{Fujinami2013}. However, the testing protocol for a handheld version of the ERG recording equipment has not been standardized to enable comparison of results across primary eye care settings. Additionally, the non-handheld units are available only in specialized retinal clinics. 
\\
Although the clinical phenotypes of various IRDs are well-documented, disease progression and prognosis can be more difficult to predict due to the heterogenous nature of the condition. The lack of clarity, uncertainty and “vague” language used by clinicians regarding the unpredictable progression of a patient’s disease and the potential timing of vision loss, can be a source of anxiety for patients with IRDs \cite{D'Amanda2020}. Having a clearer picture of the risk of progression and what to expect would not only provide peace of mind to patients, but also prove useful in helping patients plan their futures, from accessing low vision services, to gene therapies if applicable. 
For example, it is currently not possible to accurately predict individual disease progression - such a task would require analysis of large amounts of longitudinal data.
Artificial intelligence (AI) has emerged as a tool to aid practitioners in finding patterns in large amounts of varied data, thus showing promise in potential clinical applications.
\\
AI has already demonstrated significant potential in segmentation of IRD images and detection of IRD from various imaging modalities. Machine learning and deep learning algorithms, particularly convolutional neural networks, have been employed to analyze retinal images with high accuracy, aiding in the segmentation of features from images of various IRDs, and subsequent IRD detection and classification. For instance, AI models have been trained to detect features of RP and other IRDs from OCT \cite{Yang2011, Wang2018} and en face images \cite{Brancati2018, Charng2020, Zhao2022}, providing rapid and reliable assessments that can augment clinical decision-making. This review will delve into these techniques in greater detail.  Despite these advancements, challenges remain in implementation of AI into the clinic. These include the need for large, well-annotated datasets, which rely on expert human marking to accurately label features for algorithm training — a process that demands time, expertise and the availability of a ground truth such as histology. Additionally, there is a need for models that can generalize across diverse populations and the integration of explainable AI to ensure transparency and trust in clinical settings. Table \ref{Summary table} illustrates an overview of AI techniques discussed in this review, and their applications in IRDs.
\\
The prevalence and impact of IRDs necessitate comprehensive research to understand and mitigate these conditions. Although there are existing reviews on AI applications in eye care \cite{Hogarty2019, Benet2022, Han2022a, Charng2023}, there is a significant gap in studies that focus specifically on AI's role in IRDs. Many existing surveys cover general applications of AI in ophthalmology, such as diabetic retinopathy and age-related macular degeneration, but they do not delve deeply into the unique challenges and opportunities presented by IRDs.
\\
This review aims to fill this gap by systematically reviewing AI techniques used in diagnosing and managing IRDs, providing an analysis that can guide future research and clinical practice. By consolidating current knowledge, identifying gaps, and offering a structured pathway for advancing clinical applications, this review addresses the specific needs of IRD research and aims to facilitate the development of robust, interpretable AI models tailored to these complex genetic conditions.
\\
The structure of the remaining review is as follows:  \\
Section 2 provides an overview of AI techniques, emphasizing those most prevalent in eye care. \\
Section 3 delves into AI techniques applied within the field of IRDs. This section begins by examining traditional AI techniques and their applications in segmentation, classification, and predictive modelling of visual function in IRDs. It then focuses on neural networks, particularly convolutional neural networks (CNNs), which are prominent in the literature on this subject. Table \ref{Summary table} summarizes the AI techniques discussed in this section.\\
Section 4 explores both neural network and non-neural network-based AI techniques utilized in the study of other eye conditions, providing insights into potential IRD applications. \\
Section 5 concludes the review by identifying gaps in the current literature.
\\



\begin{table}
\resizebox{\textwidth}{!}{%
  \begin{tabular}{|l|l|l|l|}
\hline
 \textbf{Task} & \textbf{Model} & \textbf{Application} & \textbf{ML/DL} \\
\hline
Segmentation & \makecell[l]{Canny edge detector \cite{Yang2011}\\ Decision trees\cite{Brancati2018} \\ Random forests \cite{Wang2018}} & \makecell[l]{Segment retinal layers and \\ structures in images of IRDs} & ML \\ 
\cline{2-4}
& \makecell[l]{U-Net \cite{Ronneberger2015} \\ Modified U-Net \cite{Charng2020, Zhao2022, Kugelman2020, Charng2024, Eckardt2024}} & \makecell[l]{Used widely in medical \\ image segmentation} & DL (CNN) \\
\cline{2-4}
& \makecell[l]{Sliding windows CNN \cite{Wang2021}} & \makecell[l]{Segment retinal layers \\ affected in IRD} & DL (CNN) \\
\hline
Classification & \makecell[l]{Ensemble model \cite{Glinton2022}: \\ Support vector machine, \\ AdaBoost, \\ Logistic regression} & \makecell[l]{Classify various subtypes \\ of IRD} & ML \\
\cline{2-4}
& \makecell[l]{Inception V3 \cite{Chen2021}\\ Inception ResNet V2 \cite{Chen2021}\\ XCeption \cite{Chen2021}\\ VGG-16 \cite{Masumoto2019} \\ VGG-19 \cite{Shah2020}} & \makecell[l]{Classify various subtypes \\ of IRD} & DL (CNN) \\
\hline
Prediction & Random forests \cite{Rohm2018} & Predict progression of IRD & ML \\
\cline{2-4}
& \makecell[l]{AlexNet \cite{Liu2023}\\ DenseNet-161 \cite{Liu2023}\\ ResNet-50 \cite{Liu2023}\\ ResNet-152 
\cite{Liu2023}\\ EfficientNetB0 \cite{Nagasato2023}\\ Inception V3 \cite{Nagasato2023} \\ VGG-16 \cite{Nagasato2023}\\ U-Net \cite{Pontikos2022}} & \makecell[l]{Predict progression of IRD - \\ structural and functional} & DL (CNN)\\
\hline
\end{tabular}}
\caption{Overview of AI methods applied in inherited retinal disease (IRD) research. Segmentation models like U-Net are used for retinal image segmentation, while classification models such as Inception V3 and VGG-16 are employed to classify different types of IRDs from retinal images. Prediction models including AlexNet and DenseNet-161 are used for forecasting disease progression. Machine learning (ML); deep learning (DL); convolutional neural network (CNN).}
\label{Summary table}
\end{table}

\section{Background}
\subsection{Artificial Intelligence}
The field of AI has risen in popularity in recent years. Fundamentally, AI is driven by machine learning (ML) to infer patterns and relationships from training data. This allows AI-equipped computer systems to perform complex problem-solving tasks. Effective training of AI models allows them to make predictions on new unseen data - based on these learned representations \cite{Jordan2015}.
\subsubsection{Machine learning}
The field of AI encompasses a range of ML techniques (or models) that vary in architecture, size and algorithmic complexity. 
\\
Decision trees are intuitive models that recursively partition data into subsets based on features, leading to a final decision or prediction at the \textit{leaf} nodes. They are valued for their interpretability and applicability in various domains. Support Vector Machines (SVMs) are another powerful ML tool, focusing on finding the optimal hyperplane that separates different classes of data with the largest possible margin \cite{Jun2021}. SVMs exploit various kernel functions \cite{Pisner2020} to handle linear and non-linear classification tasks. ML models also typically include data pre-processing and transformation layers to supplement and improve learning processes. Such transformations are often utilized for data with distinguishable patterns and less noise \cite{Pasupa2016}. 
\\
\subsubsection{Deep learning}
\begin{figure}[h]
    \centering
    \includegraphics[width=1\linewidth]{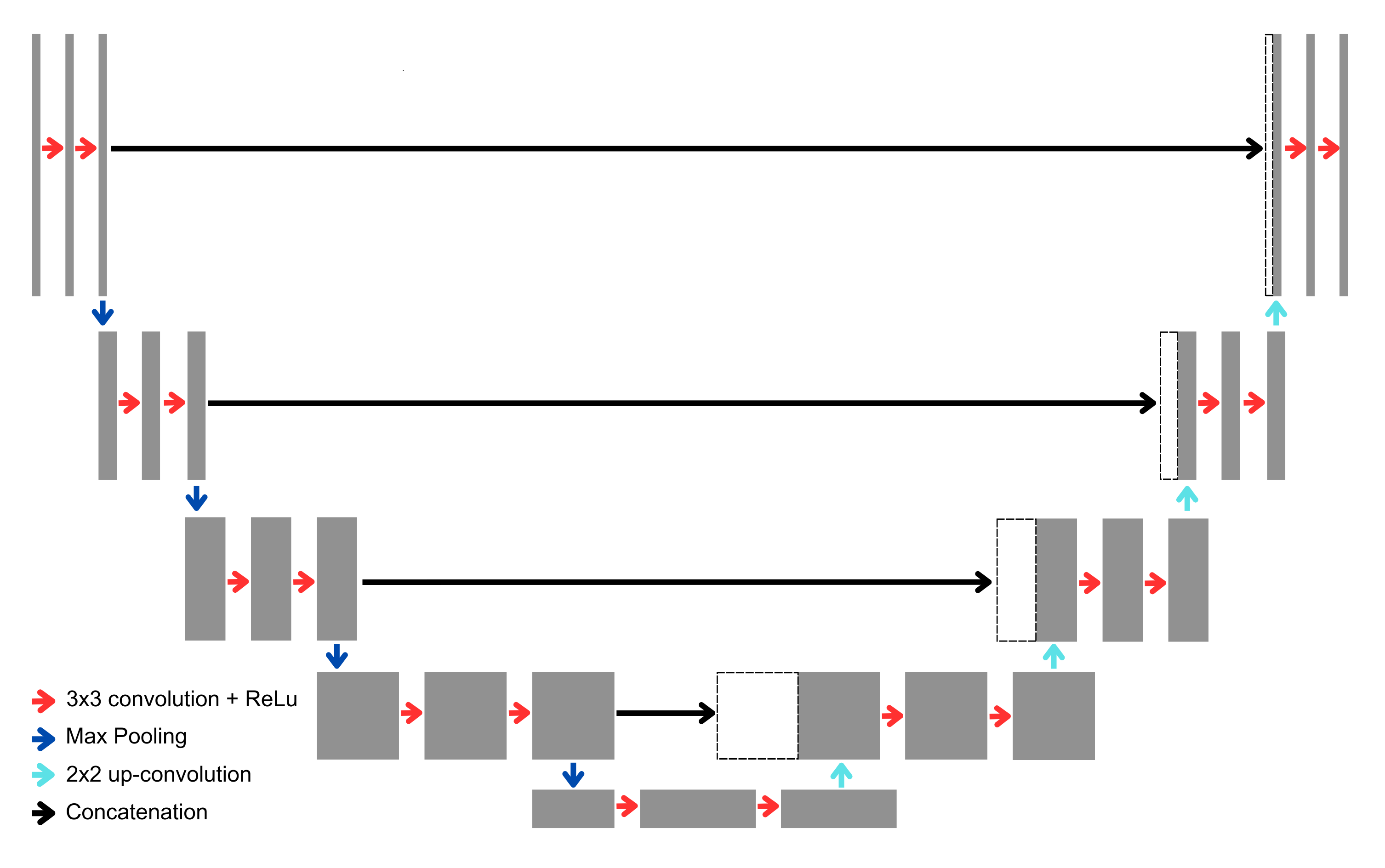}
    \caption{Schematic diagram illustrating the U-Net architecture. The network consists of an encoder path with convolutional and max-pooling layers, and a decoder path with up-convolutions. Skip connections between corresponding layers in the encoder and decoder allow for the combination of detailed spatial information and contextual features, enhancing segmentation accuracy. White dotted boxes represent concatenation of previous layers in the encoder path with up-convoluted layers at various stages of the decoder path.}
    \label{unet}
\end{figure}

Conversely, modern applications of AI often employ deep neural networks in learning processes typically referred to as deep learning (DL), which is an evolutionary sub-class of ML. DL refers to neural networks with many layers and the ability to learn intricate patterns and complex relationships from large amounts of data \cite{LeCun2015}. DL leverages backpropagation and gradient descent functions, which allows the network to learn from itself and optimize weight functions \cite{Hochreiter2001, Rumelhart1986}. The `depth' of DL alludes to the architectural complexity of these algorithms, which are often based on neural networks as described below. This focus on deep learning is particularly relevant for IRDs due to its ability to capture intricate patterns in large datasets, which is essential for accurately classifying and potentially predicting disease progression.
\\
\\
\underline{Multilayer perceptrons}
\\
One of the simplest forms of DL models is the multilayer perceptron (MLP), which is a type of neural network. An MLP consists of multiple layers of nodes, each layer fully connected to the next one \cite{LeCun2015}. The layers include an input layer, one or more hidden layers, and an output layer. Each node, or neuron, in a layer applies a weighted sum of its inputs and passes the result through an activation function to introduce non-linearity. MLPs are capable of learning complex patterns by adjusting the weights during training through techniques such as backpropagation and gradient descent \cite{Ruck1990}. Despite their simplicity, the seminal MLP is powerful and serves as the foundation for more advanced neural network architectures.
\\
\\
\underline{Transformers}
\\
Transformers represent a significant advancement in the field of DL, particularly in natural language processing but increasingly in other domains, including computer vision. Transformers are neural networks that use self-attention mechanisms to weigh the importance of different input elements dynamically \cite{Vaswani2017}. This architecture allows transformers to process sequences of data, such as text or time-series data \cite{Cholakov2021}. The self-attention mechanism enables transformers to capture long-range dependencies and interactions within the data, leading to superior performance in tasks like language translation \cite{Liu2019a}, text generation \cite{Koncel2022}, and even image recognition when adapted for vision tasks \cite{Meng2022}. Exploiting sequential data structures has also proved beneficial for transformer-based genome analysis and sequence labelling \cite{Choi2023}.
\\
\\
\underline{Convolutional neural networks}
\\
Convolutional neural networks (CNNs) are a type of neural network inspired by the mammalian visual cortex \cite{Serre2007}. They use the concept of receptive fields in order to extract features from input images \cite{Luo2016}. CNNs are typically comprised of a wide range of functional `layers'. For example, convolution and pooling layers extract features, whereas connected layers map these features to output \cite{Chua1993}. Neural networks represent a series of repeating layers which enhance recognition of complex features and relations. Many CNN models have been proposed in the literature, with varying architectures differing in size and number of repetitions within layers \cite{Krizhevsky2012, LeCun1998, Szegedy2016}. For example, U-Net \cite{Ronneberger2015}, widely used in medical imaging, is composed of 23 layers and follows a symmetric encoder-decoder structure. The encoding path consists of repeated convolutional layers and max-pooling operations, which reduce the spatial dimensions while increasing the depth of feature maps. The decoding path uses up-convolutions (or transposed convolutions) to gradually restore the original spatial resolution. U-Net also incorporates skip connections between corresponding layers in the encoder and decoder paths, allowing for the combination of high-resolution features from early layers with the upsampled features. This helps to retain spatial details, making U-Net particularly effective for tasks like image segmentation, where precise localization is critical \cite{Ronneberger2015}. (Figure \ref{unet}). It comprises of an encoder, which extracts features and reduces spatial resolution, and a decoder, which uses up-convolution operations to upsample the feature map and recover spatial information. 

\begin{figure}[h]
    \centering
    \includegraphics[width=1\linewidth]{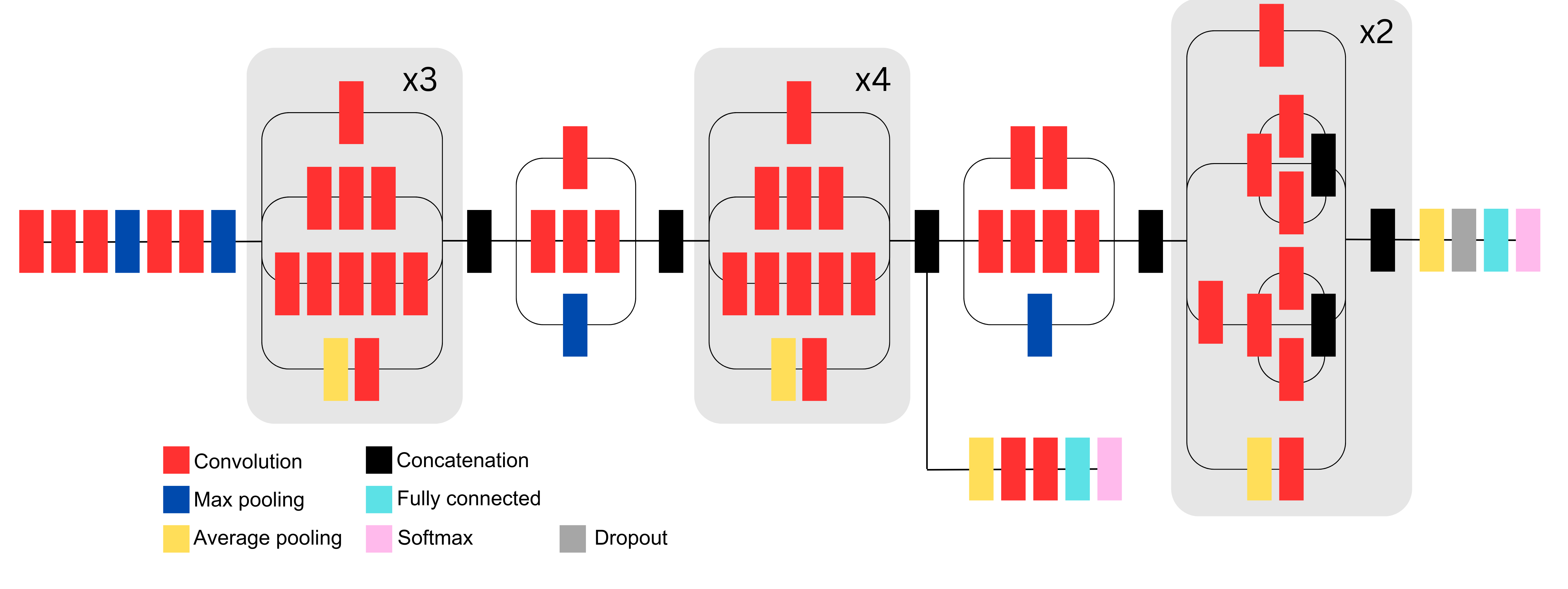}
    \caption{Schematic diagram illustrating the architecture of Inception V3. The network employs Inception modules with parallel convolutions using various filter sizes, enabling the extraction of multi-scale features. Pooling layers reduce the dimensionality of feature maps, while batch normalization and auxiliary classifiers improve training stability and classification accuracy.}
    \label{inception}
\end{figure}

In comparison, the Inception V3 model comprises 42 layers. It utilizes factorized convolutions and smaller filters to reduce computational complexity while maintaining high accuracy. The network is composed of multiple Inception modules, each performing convolutions with different filter sizes (e.g., 1x1, 3x3, 5x5) in parallel, allowing it to capture diverse features across multiple scales. The architecture also includes pooling layers to reduce dimensionality, along with techniques such as batch normalization, auxiliary classifiers, and label smoothing to enhance training stability and overall performance \cite{Szegedy2015}. (Figure \ref{inception}). 
\\ Ensemble learning allows for the merging of predictions from multiple ML models to improve performances \cite{Webb2004}.

\subsection{Artificial intelligence in eye health}

The utility of AI in eye care has been increasingly explored in recent years. AI aims to enhance disease detection and diagnosis, support clinical-decision making, screening patients, and advancing teleophthalmology \cite{Han2022b, Gunasekeran2022}. Techniques ranging from more traditional, ML-based random forest and SVM models to deep learning models like CNNs and transformers, display a promising trend toward improving image-based analysis for segmentation, classification, and risk prediction in the ophthalmology domain.
\\
Image segmentation is a critical step in data interpretation, extracting meaningful components from inputs to isolate relevant areas based on specific attributes \cite{Ghosh2019}. This process is crucial for the retina, which consists of multiple layers and features requiring accurate delineation for analysis, classification and image registration. Segmentation has been applied to OCT images \cite{Wang2022}, fundus photos \cite{Chen2021, Liefers2020} and FAF images. Following segmentation, AI models can be leveraged to classify input samples, labelling them with specific diseases or disease stages, depending on how the model was trained. AI models have successfully detected various ocular conditions such as glaucoma \cite{Mai2024} and RP \cite{Chen2021, Masumoto2019}, and classified conditions such as AMD \cite{Elsharkawy2024}, glaucoma \cite{Phene2019}, Stargardt disease \cite{Shah2020, Miere2021} and various retinal conditions \cite{Miere2020}. 
\\
The selection of AI models for data analysis depends on the task at hand. For instance, SVM is suitable for binary classification, whereas logistic regression and deep learning models are better at handling multi-class problems \cite{Hosmer2013}. The size of the dataset also influences the choice of AI model. Deep learning models perform better when exposed to larger datasets, while shallow ML models are more accurate with smaller, labelled datasets \cite{Bengio2009}.
\\
Another crucial consideration is explainable AI (XAI).
AI models have often been described as ‘black-box’ systems, whereby outputs are generated without knowing what steps the model takes \cite{Duran2021}. Explainable AI practices allow for examinations of DL and ML decision-making processes \cite{Lent2004}, thereby increasing transparency and reliability. Shallow ML algorithms such as logistic regression and decision trees are inherently explainable, as each step in the model is interpretable. In contrast, deep learning models such as CNNs typically rely on saliency maps to extract relevant learned features \cite{Chaddad2023}. Some of the most commonly used XAI techniques in relation to CNNs include gradient-weighted class activation mapping (Grad-CAM) \cite{Vinogradova2020}, which produces a weighted map of salient features.
\\
XAI holds particular significance in healthcare by improving interpretability and fostering trust among practitioners and patients in AI models \cite{Samek2019}. It also carries ethical and legal implications when applied in clinical practice \cite{Goodman2017, Hacker2020}. Integrating XAI into clinical ophthalmology has shown to enhance diagnostic accuracy without increasing the time required for diagnosis \cite{Xu2021}, and is generally well-received by clinicians compared to systems lacking XAI assistance \cite{Singh2021}.

\section{AI techniques in inherited retinal diseases}

\subsection{Machine learning}

\subsubsection{Segmentation}
IRDs manifest a spectrum of clinical signs, highlighting the necessity for accurate segmentation crucial for precise diagnosis. 
Detailed segmentation of OCT images is crucial for clinicians to analyze affected retinal layers, which is pivotal for the diagnosis and management of IRDs \cite{Gersch2022}. Similarly,  accurate segmentation of en face images, particularly FAF, is clinically significant, as key features in IRDs - such as pigmentation or autofluorescence changes - can vary in size, number, and location. The ability to measure and track these changes over time offers valuable clinical insights. In the context of IRDs, segmentation refers to the algorithm’s ability to identify and outline characteristic signs or areas of interest in clinical data, including flecks, areas of atrophy, and pigmentation. The ability to produce measurements from these outlined areas allows for further processing, such as subsequent detection, classification and image registration. However, current approaches to image segmentation face challenges, including variability in image quality, differences in anatomical structures among patients, and the need for extensive training datasets to improve accuracy.
\\
Several studies have employed ML techniques to segment relevant features in IRDs. For instance, a method combining Canny edge detection and shortest-path search, accurately detected inner and outer segment contours from OCT images in patients with RP, achieving segmentation results comparable to manual methods \cite{Yang2011}. Moreover, ensemble classifiers based on decision trees were used to segment RP pigment signs from fundus images. Pre-processing and watershed transformation were applied, with findings indicating comparable discriminatory power and computational efficiency between random forests and AdaBoost models \cite{Brancati2018}. In choroideremia, a random forest model effectively detected preserved ellipsoid zones on OCT images, demonstrating substantial accuracy with a Jaccard similarity index of 0.845 compared to expert graders \cite{Wang2018}. The ellipsoid zone is a layer of the retina particularly important for vision, and disruption to this layer indicates visual dysfunction. The ability to detect areas of preserved ellipsoid zone has clinical implications for disease severity and visual function. The model proposed by Wang et al.

\subsubsection{Classification}
Classification for IRD refers to an AI model identifying a particular genotype or disease stage. A hierarchical soft-voting ensemble model was deployed by Glinton et al. \cite{Glinton2022}, combining: (\textit{i}) SVM, (\textit{ii}) AdaBoost, (\textit{iii}) decision trees and, (\textit{iv}) logistic regression to classify ERG phenotypes in patients with Stargardt disease. For patients with this disease, ERG results vary depending on which types of photoreceptor cells (i.e., rods and/or cones) are affected.  The ensemble model successfully differentiated between normal tracings and generalized photoreceptor dysfunction, achieving high R-squared values of 0.967 and 0.938 respectively compared to ground truths. However, classification of generalized cone dysfunction proved challenging due to the small sample size in this subgroup (44 patients), yielding a lower R-squared value of 0.393. 

An ensemble SVM was utilized to classify whether childrens' pupil reactions aligned with 'disease (RP)' or 'disease not detected' \cite{Smita2023}. This study used data from a binocular pupillometer, which measures pupil diameter, constriction/dilation and delay to infer retinal sensitivity. The ensemble SVM method achieved 86.55\% accuracy and 92.7\% sensitivity when compared with true diagnosis, however its clinical applicability is limited as pupillometry is mainly used in research settings rather than routine optometry or ophthalmology practices.

\subsubsection{Prediction of visual function}
For predicting visual function, random forests were utilized to estimate retinal sensitivity measured by perimetry in RP and Leber congenital amaurosis \cite{Sumaroka2019}. Additionally, random forest regression was applied to OCT images of patients with recessive Stargardt disease to predict retinal sensitivity using microperimetry \cite{Muller2021}. 
While machine learning plays a significant role in IRDs, the literature emphasises a shift towards deep learning-based methodologies, particularly when handling large image datasets.

\subsection{Deep Learning}

\subsubsection{Convolutional neural networks}
\begin{figure}[h]
    \centering
    \includegraphics[width=0.8\linewidth]{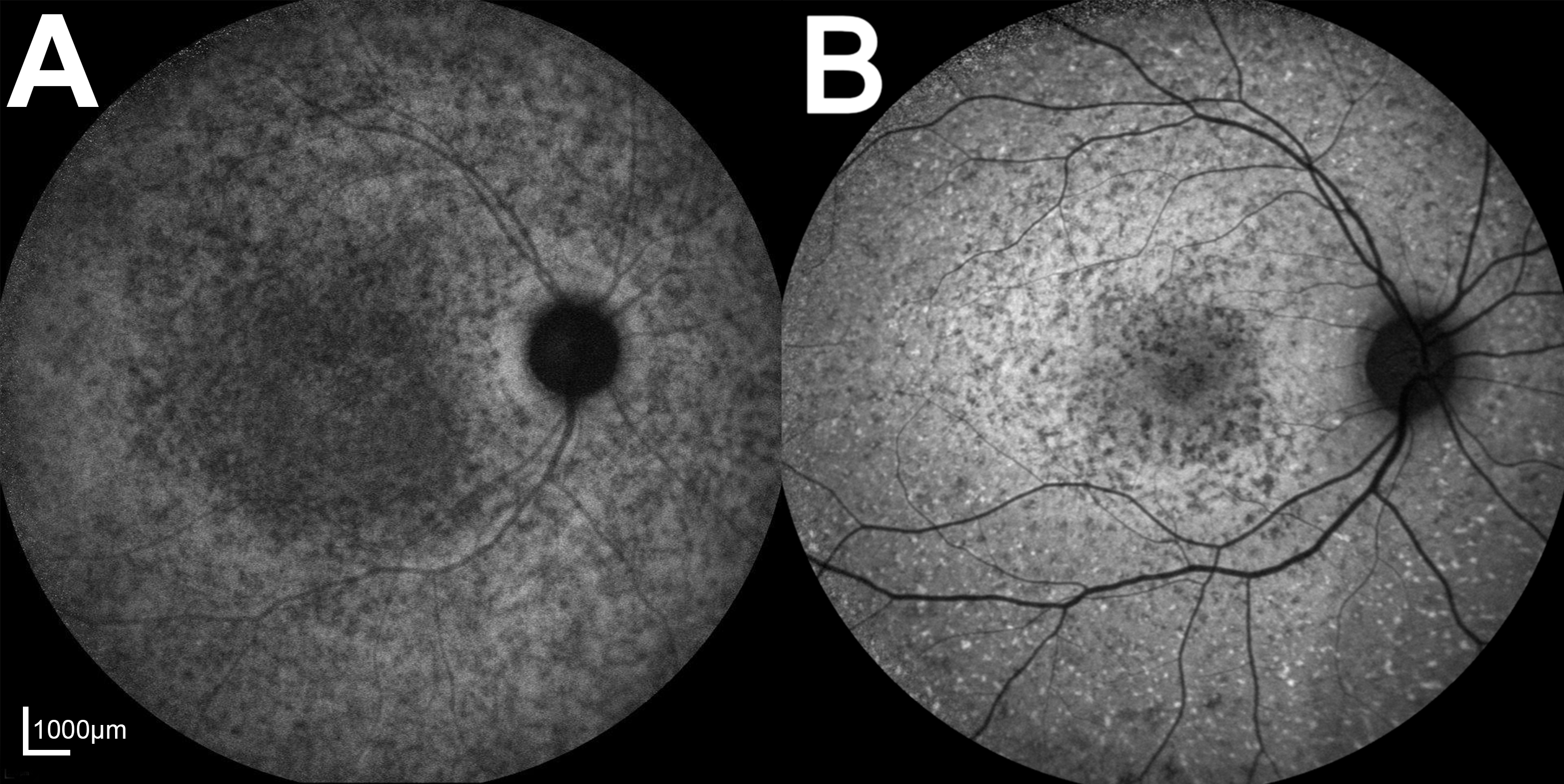}
    \caption{Hyper- and hypoautofluorescent flecks in Stargardt disease as visualized by (A) near-infrared autofluorescence and (B) short wave autofluorescence.}
    \label{stargardtAF.png}
\end{figure}

\begin{figure}[h]
    \centering
    \includegraphics[width=0.8\linewidth]{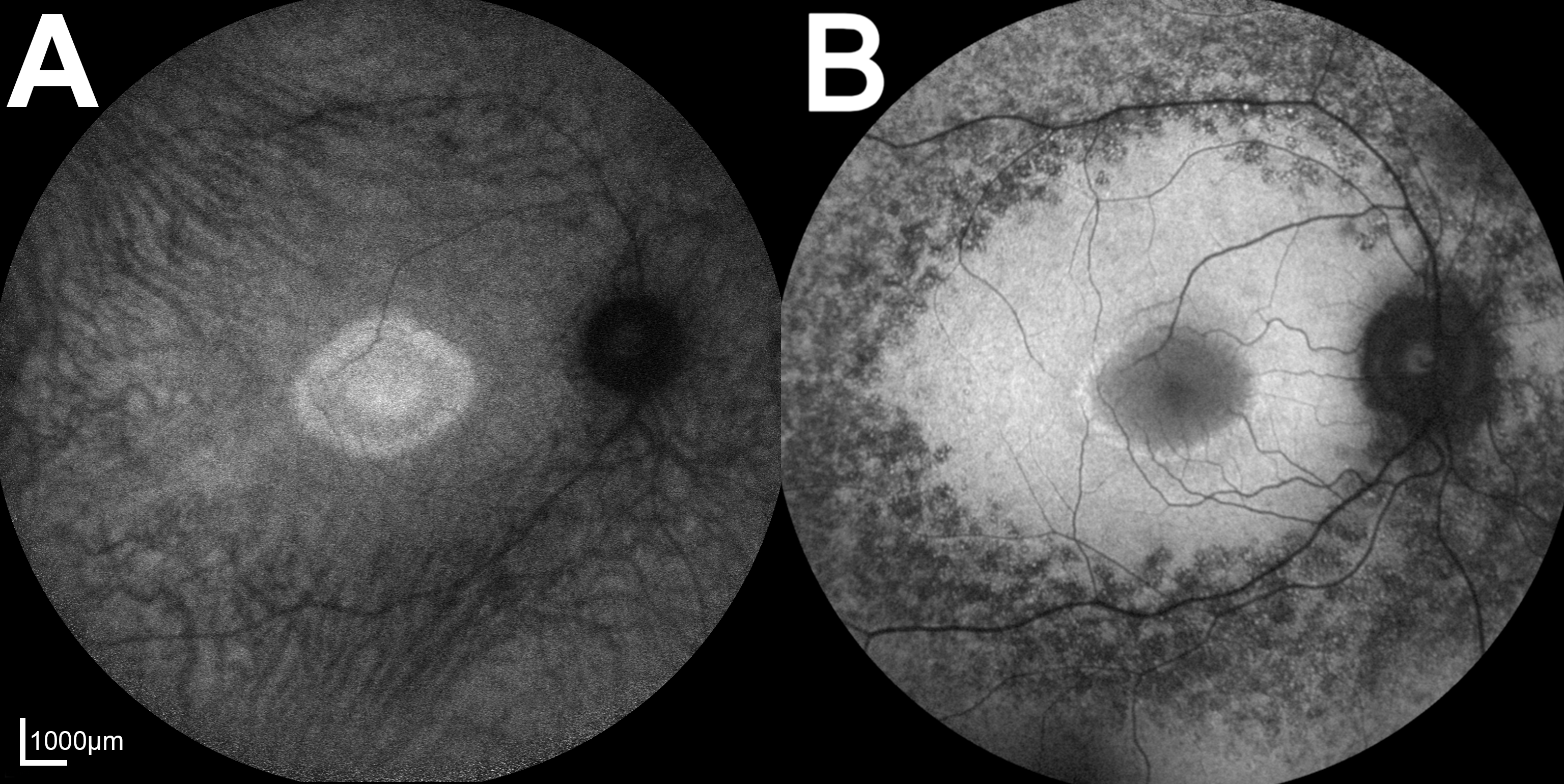}
    \caption{Hyperautofluorescent ring in retinitis pigmentosa visualized by (A) near-infrared autofluorescence and (B) short wave autofluorescence. Abnormal pigmentation is more easily visualized with short wave autofluorescence.}
    \label{RPAF.png}
\end{figure}

\underline{Segmentation}
\\
CNNs have been pivotal in segmenting critical features of various IRDs, including Stargardt disease. A hallmark feature of Stargardt disease is alterations in FAF, such as hyperautofluorescent retinal flecks indicating buildup of abnormal melanin or lipofuscin, or areas of decreased autofluorescence indicating retinal layer atrophy (Figure \ref{stargardtAF.png}). Accurate detection and quantification of hyperautofluorescent flecks are clinically significant for monitoring disease progression \cite{Solberg2019}. A modified U-Net architecture incorporating a ResNet encoder-decoder has successfully segmented hyperautofluorescent flecks \cite{Charng2020}. This adaptation replaced the traditional U-Net encoder with ResNet-34. As fleck appearance can vary, this model was tested on FAF images displaying a wide variety of presentations of Stargardt disease, and demonstrated higher Dice scores when estimating fleck number and area in discrete flecks compared to diffuse speckled patterns. This study utilized subjects with confirmed ABCA4 genetic diagnosis, thus strengthening the validity of ground truth images, however contrast limited adapted histogram equalisation (CLAHE) transformation was not used consistently across images for model segmentation compared to manual segmentation.
A similar architecture was proposed by Zhao et al. \cite{Zhao2022} more recently, replacing the encoder portion of a base U-Net with a ResNet50 model in this case in order to segment various hypoautofluorescent lesions in Stargardt disease. Detection and quantification of hypoautofluorescent lesions is of clinical importance, as changes to these lesions over time can indicate disease progression. This model did not utilize CLAHE transformation during data augmentation, and achieved an intra-class correlation coefficient of 0.997 for lesions with definitely decreased autofluorescence compared to ground truths. 
\\
Furthermore, a CNN based on U-Net was utilized to segment inner and outer retinal boundaries in OCT images from patients with Stargardt disease \cite{Kugelman2020}. This CNN comprised four pooling layers and incorporated squeeze and excitation blocks.
\\
RP also demonstrates characteristic clinical signs, and CNNs have effectively segmented these features. A significant number of RP patients exhibit a hyperautofluorescent ring surrounding the macula, visible through FAF imaging \cite{Dysli2018}, as well as abnormal pigment spots across the retina, visible via fundus photography (Figure \ref{RPAF.png}) \cite{Schuerch2016}. Accurate detection and segmentation of these signs are crucial for staging and monitoring disease progression \cite{Hamel2006, Lima2009}. An architecture combining U-Net++ with an Inception-ResNet-V2 encoder was used to segment the hyperautofluorescent ring area in FAF images for longitudinal analysis \cite{Charng2024}. Pigment signs in RP were segmented using a CNN named Retinitis Pigmentosa Segmentation Network (RPS-Net) \cite{Arsalan2020}. Unlike traditional image classification models, RPS-Net lacks a fully connected layer and minimizes feature loss through deep-feature concatenation and fewer convolutions. The model achieved an average area under the receiving operator characteristic (AUROC) of 0.807 compared to expert grading and demonstrated the ability to detect barely visible pigment signs.
\\
Another hybrid model was utilized to segment OCT images in RP \cite{Wang2021}, and a further study by the same authors used the same model to obtain three-dimensional outer segment measurements in patients with RPGR-associated RP \cite{Wang2023}. Wang et al. \cite{Wang2021} compared the accuracy of U-Net, a sliding window CNN, and a hybrid model in segmenting OCT images in RP, and found the hybrid model to slightly improve mean accuracy compared to individual models (91.5\% hybrid accuracy compared to 90.8\% and 90.7\% for U-Net and sliding-window respectively as demonstrated by Bland-Altman analysis). The authors’ 2023 study also utilized longitudinal data, albeit from a small cohort of 34 patients, which could be useful in prediction of disease progression \cite{Wang2023}. 
\\
A CNN was employed to segment areas of preserved ellipsoid zone on OCT scans in RP and choroideremia patients \cite{Camino2018}. Utilizing 61,200 OCT patches per IRD from 10 eyes with RP and 10 with choroideremia, the CNN comprised three convolutional layers with Rectified Linear Unit (ReLU) activation, three pooling layers, and two fully connected layers. The model demonstrated strong similarity with manual grading, achieving Jaccard indices of 0.894 for RP and 0.912 for choroideremia. However, limitations included a small training dataset and lack of validation on external data.
\\
 A recent study deployed a CNN based the popular U-Net architecture to segment retinal layers on OCT scans of patients with various IRDs \cite{Eckardt2024}, achieving a Dice score of 98.7\%, however the training dataset was limited to just 16 IRD patients.
 \\
Although adaptive optics scanning laser ophthalmoscopy (AOSLO) is not widely used in clinical practice, it provides high-resolution images of the retina at a cellular level. Convolutional neural networks (CNNs) have been employed to segment cones in conditions such as achromatopsia and choroideremia. For instance, a modified late fusion architecture using probability maps was utilized to infer cone locations in achromatopsia \cite{Cunefare2018}. The same CNN architecture was also applied to localize cones in AOSLO images of eyes with choroideremia \cite{Morgan2020}, demonstrating the model’s generalizability across different IRDs. CNNs are chosen for these tasks due to their ability to effectively learn and extract complex features from high-dimensional image data, making them well-suited for analyzing detailed retinal structures.
\\
Overall, CNNs successfully segment pertinent features of IRDs, however the literature so far seems to be limited by small training datasets and lack of external validation. Furthermore, there is also a lack of segmentation studies that also incorporate explainable AI techniques.
\\
\underline{Classification}
\\
CNNs have been effectively deployed for detection and classification of various IRDs across a range of imaging modalities. Chen et al. \cite{Chen2021} developed three CNNs to detect the presence of RP from color fundus photos. These DL models were based on Inception V3, Inception Resnet V2 and Xception architectures, pre-trained on the ImageNet dataset. The Xception model achieved the highest area under the receiver operating characteristic curve (AUROC) at 80\%, likely due to its additional residual connection and use of depthwise separable convolutions compared to Inception models \cite{Chollet2017}. Fine-tuning the number of convolution layers further increased the AUROC to 99.46\%. This model matched retinal and IRD specialists in accuracy, precision, and sensitivity. Grad-CAM \cite{Vinogradova2020} showed that contrast between macular and peripheral retinal regions is significant in early RP identification. However, the model's performance is limited by its small, homogeneous sample and lack of external validation.
\\
ResNet 101 has been successful in multi-class classification of Stargardt disease, RP, Best vitelliform macular dystrophy (BD), and normal FAF images \cite{Miere2020}. Pre-trained on the ImageNet dataset, ResNet 101 achieved AUROCs of 0.998, 0.999, and 0.995 for detecting Stargardt disease, RP, and BD, respectively. Integrated gradient visualization highlighted clinically relevant areas for disease classification, such as autofluorescence around the fovea in RP and vitelliform deposits in BD. However, the authors did not specify disease staging or severity, as later-stage IRDs are generally easier to distinguish. This group also used ResNet50V2 to differentiate FAF images of ABCA4-related Stargardt disease from pseudo-Stargardt pattern dystrophy caused by PRPH2/RDS variants \cite{Miere2021}. The model achieved an AUROC of 0.89 and outperformed retinal specialists in classification accuracy.
\\
Shah et al. \cite{Shah2020} used visual geometry group-19 (VGG-19), as well as a proposed CNN, to classify OCT images as normal, mild, or severe Stargardt disease. The VGG-19 model was pre-trained on ImageNet, with only the end layers retrained. Their proposed CNN, based on LeNet, included two convolutional layers with ReLU activation, two pooling layers, and three fully connected layers, using 256x32-pixel columns extracted from OCT images as input. Batch normalization was applied before ReLU activation to reduce overfitting. The VGG-19 model demonstrated sensitivities of 96.0\% and 99.5\% for mild and severe Stargardt disease classification, respectively, and a specificity of 98.0\% with a Jaccard score of 0.990 for overall binary classification of Stargardt disease vs. normal. In comparison, their proposed CNN achieved AUROCs of 0.838 for mild Stargardt disease and 0.937 for severe disease. The authors noted that the under-representation of mild phenotypes in the training data (80 mild scans versus 120 severe scans) could explain the poorer identification. Explainable AI techniques may also help clarify why mild Stargardt disease is more difficult to classify for their LeNet-based CNN model as compared to the VGG-19 counterpart.
\\
Visual geometry group-16 (VGG-16) was used to classify AF and pseudocolor ultra-wide retinal images into normal or RP \cite{Masumoto2019}.
The VGG-16 model consists of five blocks with convolution layers, max-pooling layers and two fully connected layers, demonstrating excellent performance, reporting AUROCs of 0.998 and 1.00 for pseudocolor and AF images, respectively \cite{Masumoto2019}. Heatmaps indicated that key clinical signs in RP, such as bone spicule pigmentation, were areas of interest. Despite the model's accuracy and clinical parallels, it did not distinguish between different RP stages.
\\
CNNs have demonstrated accuracy in detecting IRDs, but few studies have explored using CNNs to classify disease severity, which may have greater clinical relevance. More research is needed to investigate CNNs' potential in staging and classification given the broad diversity in IRDs.
\\
\underline{Prediction of visual function 
}
\\
From a patient perspective, visual function is one of the most clinically relevant outcomes to measure and predict, as it directly impacts daily life and wellbeing \cite{Rudnick2019}. Studies have attempted to use CNNs to predict visual acuity (VA) and visual field (VF) sensitivity in RP. Four pre-trained CNNs (AlexNet, DenseNet-161, ResNet-50, and ResNet-152) were used to estimate VA as better or worse than 6/12 (a common cutoff for driving standards worldwide) based on OCT and/or infrared images \cite{Liu2023}. ResNet-152 performed best during 10-fold cross-validation. Grad-CAM results suggested that the model relied more on OCT data than infrared data. The AUROC for the binary classification of VA better or worse than 6/12 was 0.87 for OCT only and 0.85 for combined OCT and infrared.
\\
Another recent study used various ensemble DL models comprising EfficientNetB0, InceptionV3, and VGG-16 to estimate VA, central VF sensitivity, and mean deviation (MD, an indication of how sensitive a patient’s VF is compared to age-normative data) in patients with RP \cite{Nagasato2023}. Score-CAM-generated heatmaps indicated that the fovea, AF rings, and degeneration margins were most salient when estimating MD, consistent with ophthalmologists' assessments of RP. While this study was purposeful in using multimodal imaging and functional assessment, the sampling excluded atypical presentations of RP.  
\\
\underline{Prediction of disease progression}
\\
Aside from visual function, disease progression is another significant concern for patients with IRDs. The progression of retinal atrophy in Stargardt disease was predicted using FAF images from 206 eyes with 12-month longitudinal data \cite{Pontikos2022}. A U-Net architecture with embedded self-attention mechanisms achieved a Dice score of 0.76 in predicting atrophy after 12 months. Similarly, Veturi et al. \cite{Veturi2023} attempted to predict the rate of progression of Stargardt disease using OCT and FAF images from 237 eyes with 12-month longitudinal data. The authors used a modular series of U-Nets to produce a probability map for the predicted atrophy region, achieving Dice scores of 0.83 and 0.828 for six- and 12-month predictions, respectively. Despite these promising results, 12-month predictions may be insufficient for lifelong conditions like IRDs, limiting the ability to provide comprehensive patient education and counseling.
\\
Although significant progress has been made in investigating visual function, further research is needed to refine accuracy and test models on external datasets. While estimating current visual function is a promising start, patients may also benefit from the prediction of future visual function and potential functional improvement, areas yet to be thoroughly explored in the AI research field.
\\
\underline{Prediction of causative genes}
\\
Recent interest in gene therapy for inherited retinal diseases (IRDs) has been highlighted by the FDA's approval of Luxturna (voretigene neparvovec-rzyl), the first ocular gene therapy for RPE65-associated retinal dystrophy \cite{Rudnick2019}. Despite this progress, genetic testing can be time and financially prohibitive. CNNs have shown promise in predicting causative genes for various IRDs, aiding in genetic diagnosis, counseling, family planning, and clinical trial recruitment.
\\
One study utilized the Medic Mind platform, leveraging the Inception V3 architecture, to predict causative genes for ABCA4, RP1L1, and EYS-associated retinopathies based on OCT images \cite{Liu2023}. Mean test accuracy was 100\% for ABCA4, 78.0\% for RP1L1, 89.8\% for EYS, and 93.4\% for normal. The authors suggest that overfitting may explain the model’s lower performance for RP1L1 and EYS, attributed to the small, homogenous sample sizes of 20 and 28 patients, respectively.
\\
The Eye2Gene project \cite{Quang2023} aims to predict causative genes based on OCT and infrared/FAF imaging. This initiative developed an ensemble of fifteen Inception V3 CNNs, trained on the largest IRD dataset to date (44,817 scans from 1907 patients) and tested on four external datasets. Early results \cite{Pontikos2022} show accurate predictions for the 36 most common IRD genes, matching expert standards. The model also identified the most pertinent imaging modalities for specific genes and demonstrated improved accuracy when considering the age of first presentation and mode of inheritance.

\subsubsection{Generative adversarial networks}
A notable outcome of the Eye2Gene project was the creation of SynthEye, a generative adversarial network (GAN) designed to generate synthetic FAF images \cite{Veturi2023}. This StyleGAN2-ADA model was trained on FAF images from 36 IRD classes, producing novel, diverse images of high visual fidelity. When these synthetic images were used to train Inception V3-based models for image classification, the performance of the classifiers remained consistent, neither improved nor degraded by the addition of synthetic images.
\\
The use of GANs to generate synthetic images may help address class imbalances, although they cannot replace genuine images and the unique information they provide. Nevertheless, this study underscores the vast potential of applying GANs in the field of IRDs. 

\subsubsection{Recurrent neural networks}
While CNNs are suited for analysing spatial data, recurrent neural networks (RNNs) are more appropriate for temporal and sequential information. Given the emphasis on imaging in IRDs, RNNs are less prominent in current literature compared to CNNs. However, some studies have utilized RNNs for segmentation tasks. For example, a multidimensional RNN with convolutional layers was used to localize cones in AOSLO images of Stargardt disease \cite{Davidson2018}. The authors suggest that unlike CNNs, multidimensional RNNs provide global context at every pixel throughout classification, learning dependencies between pixels while detecting pertinent local features through convolutional layers. This network demonstrated strong performance in localizing cones in AOSLO images of Stargardt disease, achieving an average Dice score of 0.94, and was capable of detecting cones in other pathologies such as RP and achromatopsia. Despite these promising results, the clinical utility is limited since adaptive optics SLO is not commonly available in general optometry and ophthalmology practices.

\section{AI techniques in other eye conditions}
While research into AI and inherited retinal diseases (IRDs) has made significant progress, notable gaps still exist in terms of predicting visual function, risk prediction and applying explainable AI principles in ML- and DL-based models.
This section delves into AI techniques applied to more common eye conditions, aiming to explore potential applications in the field of IRDs. It should be noted that the techniques described in this section are not the only AI models utilized for their corresponding conditions, nor are they exclusively used for these conditions. They have been selected to highlight potential ways to bridge the knowledge gap in IRDs

\subsection{AI techniques in glaucoma}
Glaucoma is a condition that causes damage to the optic nerve and can lead to VF loss \cite{Harry1993}. There are various types of glaucoma, including primary open-angle glaucoma (POAG) and normal-tension glaucoma (NTG) \cite{NTG}. Ocular hypertension refers to when patients have high intraocular pressures, which is a risk factor for developing glaucoma, but have no detectable glaucomatous damage \cite{Gordon2002}. The use of AI techniques and XAI to detect structural changes is well-documented \cite{Mehta2021, Phene2019, Liu2019b, Li2019, Shyamalee2024, Guo2023}.
\\
Like IRDs, glaucoma causes progressive VF loss, making perimetry essential for assessing both conditions. While AI's role in the structural assessment of IRDs has been investigated, research into AI for VF testing in IRDs, particularly regarding VF loss progression, is lacking. Insights from glaucoma research may help bridge this gap.

\subsubsection{Machine learning in prediction of visual field progression}
Various ML techniques have been used to predict the progression of perimetry results in glaucoma patients. Kalman filtering, for instance, has been used to predict the two-year trajectory of glaucomatous VF loss in NTG patients \cite{Garcia2019}. The study found that Kalman filtering predicted future mean deviation (MD) more accurately than simple and modified regression models, especially in mild NTG cases. However, the study's homogenous sample and need for multiple baseline measurements limit generalizability. Predicting VF progression on a point-wise basis, indicating specific areas of vision loss, may be more valuable to patients than global VF loss indicators.
\\
Archetypal analysis (AA), a form of unsupervised ML, has also been used to predict VF progression \cite{Elze2015, Wang2020, Yousefi2022} and risk of glaucoma progression \cite{Singh2024} in POAG and ocular hypertensive patients with abnormal VF results. AA identified archetypes associated with structural glaucomatous changes in the retinal nerve fibre layer \cite{Elze2015}, classified various patterns of glaucomatous VF loss \cite{Yousefi2022}, and quantified central VF loss to improve prediction of future glaucomatous VF loss \cite{Wang2020}. More recently, AA identified baseline VF result archetypes that may increase the risk of VF progression and POAG onset in ocular hypertensive patients, even before detectable structural glaucomatous damage \cite{Singh2024}. This study used data collected over 20 years, demonstrating AA's potential for analysing longitudinal conditions, including IRDs.

\subsubsection{Neural networks in prediction of visual field progression}
The performance of convolutional neural networks (CNNs) compared to linear regression models in predicting VF progression in POAG patients has been assessed in \cite{Shon2022}. Here, pointwise linear regression and regression on global indices were compared to a proposed CNN. The architecture of this CNN comprised of three blocks, each containing a 3D convolution, batch normalization and ReLu activation layers, followed by two fully connected layers. VF results were converted to tensors in order to preserve spatial characteristics, and concatenated to be used as further input for an enhanced CNN model \cite{Shon2022}. The authors reported AUROC of 0.398 and 0.619 for linear regression based on global indices and pointwise linear regression respectively, and 0.864 for the the enhanced CNN model. 
\\
Similarly, CascadeNet-5, a CNN with five `cascading' convolutional layers, which copy concatenations of output tensors in the forward process \cite{Wen2019}. The authors report a lower mean absolute error when estimating 2-year VF progression at a point-wise level in glaucomatous eyes compared to linear models \cite{Wen2019}. The addition of age as a clinical factor also improved the CNN’s prediction accuracy. 
\\
A recent study \cite{Eslami2023} compared the CNN used by Wen et al. \cite{Wen2019} with a long short-term memory (LSTM) RNN used by Park et al. \cite{Park2019}. The RNN comprised a single layer of LSTM with a single-layer fully connected network. Overall, both methods had low pointwise mean absolute error, with the LSTM-based RNN demonstrating better performance in forecasting VF prediction compared to the CNN. However, the RNN required at least six consecutive VFs per patient, which may limit its clinical applicability. Furthermore, it was noted that both models underpredict worsening of VF loss \cite{Eslami2023}, thus highlighting the need for further research in this area.
\\
A generalized variational autoencoder has also been used to predict future VF patterns in patients with established glaucomatous VF losses \cite{Berchuck2019}. This model was able to generate predictive VFs and localized defects across four years. 
\\
AI has a role in predicting VF losses in glaucoma, and these techniques may be translatable to IRDs. However, these studies generally require multiple baseline and consecutive VFs, which may prove more difficult in a rarer cohort of IRD patients. There is also a notable lack of XAI integration in this area. 

\subsection{AI techniques in age-related macular degeneration}
Age-related macular degeneration (AMD) is a condition that can cause central vision loss \cite{Pondorfer2020}. Visual loss occurs in late stages through two processes: dry (atrophic) AMD involves progressive retinal layer atrophy \cite{Rickman2013}, while wet (neovascular) AMD involves new blood vessel growth that can leak fluid and blood into the retina \cite{Lim2012}. Anti-complement factor therapies have been shown to reduce the expansion rate of atrophic area \cite{Heier2023}. Anti-vascular endothelial growth factor (anti-VEGF) therapies can stabilize neovascular AMD by shutting down the growth of abnormal blood vessels in the eye, although individual responses to the therapy can vary \cite{Amoaku2015}.
\\
More research exists on AI prediction of visual function and atrophy growth rates in AMD than in IRDs, partly due to the availability of treatments such as anti-complement and anti-VEGF injections. Various AI techniques used to predict atrophy expansion in dry AMD, and VA and retinal sensitivity in neovascular AMD, may be applicable to IRDs.

\subsubsection{Machine learning in prediction of future visual function}
Most studies use VA as a measure of visual function due to its routine measurement in practice and repeatability as an indicator of macular function \cite{Aslam2014}. Five ML algorithms (AdaBoost.R2, Random Forests, Extremely Randomised Trees, least absolute shrinkage and selection operator [Lasso] and Gradient Boosting) were compared to predict VA at three and 12 month timepoints in patients with neovascular AMD following treatment with three anti-VEGF injections \cite{Rohm2018}. In addition to VA, models were also trained on OCT measurements. Of the models tested, Lasso had the lowest mean absolute error in predicting VA after 12 months. Accuracy was further improved by adding additional data from previous visits.
\\
Various decision trees have also been used in this area. AutoML and XGBoost predicted whether VA in neovascular AMD patients would be above or below driving standard cutoffs after one year of anti-VEGF treatment \cite{Abbas2022}. Both models achieved AUROCs of around 0.85. Using the What-if Tool (an XAI technique) \cite{Wexler2019}, the model's tendency to overpredict negative outcomes in Asian patients was identified and adjusted, improving accuracy. This ability to calibrate models based on explainable findings is valuable for IRDs due to their heterogeneity.

\subsubsection{Neural networks in prediction of future visual function and disease progression}
A recent study used a CNN to model the progression of geographic atrophy lesion growth, utilizing FAF and OCT images from 184 eyes, with areas of geographic atrophy manually delineated \cite{Mai2024}. Patients were followed up every three months for over 12 months. In order to predict progression at certain time points, the model estimates the time derivative for each OCT scan with a smaller neural network, which is embedded with a linear classification layer. This model achieved an R-squared value of 0.37 compared to ground truths for predicting progression rates over three years, however it was able to identify patients likely to progress quickly (that is, geographic atrophy growth of over 0.6071mm/year) with an AUROC of 0.81.
\\
Neural networks have also been utilised to predict VA for patients with neovascular AMD. The HDF-Net CNN proposed in \cite{Yeh2022} was designed for VA prediction in patients with neovascular AMD and outperfromed ResNet50 and AlexNet models when predicting the 12-month VA outcome following anti-VEGF injections \cite{Yeh2022}. The HDF-Net comprised five convolutional layers and three maximum pooling layers. There was an additional input layer following the feature extraction network, to input numerical clinical data such as patient age, gender and baseline VA. HDF-Net achieved an AUROC of 0.989, compared with 0.924 and 0.936 for ResNet50 and AlexNet respectively, thus highlighting the benefit of incorporating additional demographic information. Attention maps (i.e. heatmap) demonstrated that the model paid attention to clinically relevant features, such as preserved/disrupted ellipsoid zone, subretinal hyperreflective material, and loss of outer retinal layers.
\\
An AttentionGAN-based model was utilised to predict treatment outcomes following administration of different types of anti-VEGF agents, aflibercept and ranibizumab \cite{Moon2023}. This model achieved sensitivity and specificity of 0.857 and 0.881 respectively in predicting residual retinal fluid following treatment with aflibercept, which was superior to predictions by retinal specialists. Heatmaps were generated to highlight areas of interest, and the authors found that the heatmap images were different between aflibercept and ranibizumab. Although other patient factors must be considered when deciding on a treatment, GAN has a role in guiding the practitioner and providing data-driven predictions.

\subsection{AI techniques across multiple eye conditions}
Although previous sections of this review have explored the utility of AI in specific ocular conditions, AI can also differentiate between a range of conditions. This generalizability is particularly useful for IRDs, as this group of conditions comprises many different diseases and presentations. Most studies applying AI across multiple ocular conditions have used CNNs for their efficacy in handling large datasets and image classification \cite{Krizhevsky2012}. CNNs have been used to grade and distinguish between diabetic retinopathy (DR), AMD, glaucoma, retinal vessel occlusions \cite{Khan2023, Kayadibi2023, Skevas2024}, and more specific macular conditions such as diabetic macular oedema (DMO), drusen, and choroidal neovascularisation (CNV) \cite{Puneet2022, Subramanian2022}, however these studies often lack XAI techniques. This section focuses on research utilising XAI to diagnose various eye conditions. 
\\
The authors of \cite{Bhandari2023} created a lightweight CNN comprising only five convolution layers with increasing filter depths, to detect and classify CNV, DMO, drusen and normal fundus photos. To minimize overfitting, dropout layers randomly deactivated neurons after each pooling layer during training. The model achieved an AUROC of 0.99 for each class. The authors also utilized Local Interpretable Model-agnostic Explanations (LIME) and Shapley additive explanations (SHAP) to investigate how specific features affect model prediction, and to explore the value each pixel contributed to the prediction, respectively. This CNN and LIME/SHAP 
also identified pneumonia, tuberculosis, and kidney stones from X-rays and CT scans, demonstrating generalizability across conditions.
\\
CNV, drusen and DMO were also investigated in \cite{Sunija2021}, using OCT images. Here, the authors utilized a CNN comprising six convolution layers, with batch normalization and max pooling performed after each convolution and ReLu stage. The model was optimised using Cross-Entropy. F1 score for CNV, DMO and drusen were 0.99, 1.00 and 0.99 respectively. Grad-CAM demonstrated that the model looked at the retinal pigment epithelium and Bruch’s membrane, which are clinically significant areas for assessing these conditions. This model demonstrates the potential of a lightweight CNN to accurately diagnose retinal conditions, and for grad-CAM to generate clinically pertinent areas of interest.
\\
Similarly, a CNN based on ResNet50 was used to detect and classify CNV, DMO and drusen using OCT images \cite{Li2019}. The main difference between this CNN and ResNet50 was the introduction of dilated convolution into the convolution layers, in order to maximize image space resolution. Ensembling four architectures improved overall performance, achieving AUROC of 0.995. Occlusion testing highlighted important OCT image areas for classification, corresponding to clinically significant pathology regions.

\section{Conclusion}
The current literature has been successful in segmenting and classifying IRDs using various AI techniques, however more research is required to predict visual function and future progression. Most research in the field of AI and IRDs has been conducted on imaging data, such as OCT and fundus photos, hence it may prove beneficial to further investigate VA, VF, and even ERG data. It may also be of use to incorporate patient data such as medical history, demographics, and genetic information for a more holistic approach. AI can process large amounts of data and infer patterns, which provides opportunity for disease modelling and prediction, which is not currently available for IRDs. 
\\
This can aid in providing personalized care and assisting with clinical decision making. Techniques used to investigate other ocular conditions may be translatable to IRDs.
The role of XAI is emerging as crucial to not only strengthen clinician and patient trust in the models, but also to facilitate troubleshooting of the model, enhance accountability and alleviate ethical and legal concerns. Patient and practitioner feedback and acceptance must also be assessed prior to implementing any AI tools in practice. 
\\
There is much potential for the use of XAI techniques in IRDs. Although there is still considerable research to be done, it paves the way for greater data-driven patient care.

\bibliographystyle{elsarticle-num}
\bibliography{AI-techniques-in-IRDs}

\end{document}